Secondary use of employee COVID-19 symptom reporting as syndromic surveillance as an early warning signal of future hospitalizations

Steven Horng, MD MMSc[1,2] *

Ashley O'Donoghue, PhD[1] *

Tenzin Dechen, MPH[1]

Matthew Rabesa, MBA[3]

Ayad Shammout[4]

Lawrence Markson, MD[4]

Venkat Jegadeesan[4]

Manu Tandon, MBA MPA[4]

Jennifer P. Stevens, MD MS[1,5]

[1]Center for Healthcare Delivery Science, Beth Israel Deaconess Medical Center, Boston, MA

[2]Department of Emergency Medicine, Beth Israel Deaconess Medical Center, Boston, MA

[3]Employee Health, Beth Israel Deaconess Medical Center, Boston, MA

[4]Information Systems, Beth Israel Deaconess Medical Center, Boston, MA

[5]Division for Pulmonary, Critical Care, and Sleep Medicine, Department of Medicine, Beth Israel Deaconess Medical Center, Boston, MA

* The two authors contributed equally as co-first authors

Corresponding author:

Steven Horng

330 Brookline Avenue

Boston, MA 02215

shorng@bidmc.harvard.edu

Word counts:

Abstract: 299 words

Main manuscript: 2709 words

Funding sources

Competing interest

None declared

**Key Points**

**Question:** Can secondary use of employee symptom attestation data be used as syndromic surveillance to forecast COVID-19 hospitalizations in the communities where the employees live?

**Findings:** In this retrospective cohort study of 10 hospitals in a hospital network, we forecasted COVID-19 hospitalizations in 7 days with a mean absolute error (MAE) of 6.9 patients and a weighted mean absolute percentage error (WMAPE) of 1.5% for the hospital network.

**Meaning:** In a novel pandemic before reliable testing is available, use of non-traditional secondary data sources can be used to forecast hospital demand.

**Abstract**

**Importance:** Alternative methods for hospital utilization forecasting, essential information in hospital crisis planning, are necessary in a novel pandemic when traditional data sources such as disease testing are limited.

**Objective:** Determine whether mandatory daily employee symptom attestation data can be used as syndromic surveillance to forecast COVID-19 hospitalizations in the communities where employees live.

**Design:** Retrospective cohort study

**Setting:** Large academic hospital network of 10 hospitals accounting for a total of 2,384 beds and 136,000 discharges in New England.

**Participants:** 6,841 employees working on-site of Hospital 1 from April 2, 2020 to November 4, 2020, who live in the 10 hospitals' service areas.

**Interventions**: Mandatory, daily employee self-reported symptoms were collected using an automated text messaging system.

**Main Outcomes and Measures:** Mean absolute error (MAE) and weighted mean absolute percentage error (WMAPE) of 7 day forecasts of daily COVID-19 hospital census at each hospital.

**Results:** 6,841 employees, with a mean age of 40.8 (SD = 13.6), 8.8 years of service (SD = 10.4), and 74.8% were female (n = 5,120), living in the 10 hospitals service areas. Our model has a MAE of 6.9 COVID-19 patients and a WMAPE of 1.5% for hospitalizations for the entire hospital network. The individual hospitals had an MAE that ranged from 0.9 to 4.5 patients (WMAPE ranged from 2.1% to 16.1%). At Hospital 1, a doubling of the number of employees reporting symptoms (which corresponds to 4 additional employees reporting symptoms at the mean for Hospital 1) is associated with a 5% increase in COVID-19 hospitalizations at Hospital 1 in 7 days (95% CI: (0.02, 0.07)).

**Conclusions and Relevance:** We found that a real-time employee health attestation tool used at a single hospital could be used to predict subsequent hospitalizations in 7 days at hospitals throughout a larger hospital network in New England.

**Introduction**

COVID-19 has created a challenge for many businesses -- to identify and limit employees sick with COVID-19 to mitigate the spread of infections. However, most organizations have limited access to surveillance testing for the disease. As a surrogate for testing, multiple businesses, including hospitals, have required that employees report any COVID-19-associated symptoms and to direct symptomatic employees to obtain follow up testing. Symptom reporting tools may also have additional secondary benefits. The Department of Veterans Affairs described how their patients' use of a symptom monitoring tool for COVID-19 improved a sense of connection,[1] suggesting such employee attestation strategies may have additional benefits to healthcare organizations.

Prior efforts to have employees self-identify have been met with varied success.[2,3] One critical weakness of symptom-only reporting is the infectivity of asymptomatic employees[4], who would only be identified with a larger surveillance testing strategy. As members of a community, however, hospital employees' routine reporting of symptoms could serve as a surrogate of symptom reporting for their community as a whole. Given that prior research has noted that community spread of COVID-19 makes up the bulk of the burden of new infections in healthcare settings [5], employee attestations of symptoms may also have a secondary use as syndromic surveillance to estimate the incidence and prevalence of infections in the communities where employees live.

We hypothesized that an employee symptom reporting tool at a single academic medical center could be used as syndromic surveillance and forecast subsequent hospital admissions for COVID-19 in the communities where employees live.

**Methods**

*Study Design*

This study is a retrospective cohort study and was determined to be exempt by our institutional review board with a waiver of informed consent.

*Study Population*

The study was performed in a large hospital system in Massachusetts, containing 10 hospitals from April 2, 2020, to November 4, 2020. The hospitals are numbered according to the number of unique employees living in that service area that are filling out the attestation during our sample period (ie: Hospital 1 has the most employees living in their service area that fill out the employee attestation form while Hospital 10 has the fewest). Hospital 1 is a tertiary, academic, teaching hospital in Boston, MA, with 719 staffed beds and 40,000 annual inpatient discharges. Hospital 2: 208 staffed beds, 13,000 annual inpatient discharges; Hospital 3: 102 staffed beds, 6,000 annual inpatient discharges; Hospital 4: 345 staffed beds, 24,000 annual inpatient discharges; Hospital 5: 164 staffed beds, 11,000 annual inpatient discharges; Hospital 6: 58 staffed beds, 2,500 annual inpatient discharges; Hospital 7: 364 staffed beds, 19,500 annual inpatient discharges; Hospital 8: 205 staffed beds, 12,000 annual inpatient discharges; Hospital 9: 140 staffed beds, 5,500 annual inpatient discharges; Hospital 10: 79 staffed beds, 2,500 annual inpatient discharges.

*Inclusion/Exclusion Criteria*

Employees were included if they were employed at the main hospital and were working on-site that day. Employees were excluded if they lived outside of any of the hospital network's service area or if they were not working on-site that day. Attending physicians were members of a separate physicians organization and were excluded.

*Data Collection Tool*

Self-reported symptoms were collected using an automated text messaging system (Figure 1). Employees received a text message each morning, asking them to complete the daily symptom monitoring assessment. Employees were first asked whether they will be working on-site that day. Only employees who report that they will be working on-site are prompted to fill out the symptom reporting form, where they are asked if they are experiencing any covid-related symptoms from a list of 12 symptoms. If yes, they are asked to report the specific symptom from the list of 12 that they are experiencing.

*Outcomes*

Our primary outcome was the daily number of COVID-19 hospitalized patients at each of the 10 hospitals in a large New England hospital network. Our secondary outcome was the weekly COVID-19 positive cases within each of the 10 hospital service areas in a large New England hospital network. This outcome has a reduced time period for analysis because of data availability from the state reporting agency.

*Study Variables*

The role of the employee (RN, Operations, Administrative Support, Research, Clinical Technician, Housestaff/Fellows/Residents/Interns, Clinical Assistants, and Other), age, years of service, sex, race (White, Black, Asian, Hispanic, Other), and zip code were collected.

*Data Processing*

We collected the number of employees reporting any covid-related symptom each day, grouped by employee home zip code. Employees' home zip codes were then matched to the service areas of hospitals within the hospital network. Employees were therefore matched to the hospital nearest to which they live, rather than the hospital at which they work. The data were smoothed using a 7 day moving average. The data were then transformed by taking the natural logarithm of the data.

*Statistical Analysis*

We report means and standard deviations of continuous variables and counts and percentages of categorical variables. We use an econometric time series forecasting model described by Dumitrescu and Hurlin[6] to model each hospital as a cross-section of a panel, and test for Granger non-causality in this heterogeneous panel. This type of model has previously been used to forecast COVID-19 cases at the country-level using Google search trends.[7] In this framework, a linear autoregressive model is used, allowing coefficients to differ across hospitals in the panel, but fixed over time. Holding the number of lags constant across hospitals, we select the optimal number of lags that minimizes the Bayesian information criterion and forecast future COVID-19 hospitalizations at each network hospital. We forecast 7 days into the future and we calculate the mean absolute error and mean absolute percentage error of our forecast to measure the accuracy of our model for the network. There are no missing days of hospital census data or employee symptom reporting data. $P \leqq 0.05$ was considered statistically significant and all tests were 2 tailed. Stata SE version 16 (StataCorp) was used for statistical analysis.

**Results**

Summary statistics about the employees are reported in Table 1. We studied a total of 6,841 employees from April 2, 2020 to November 4, 2020, with a mean age of 40.8 (SD = 13.6), 8.8 years of service (SD = 10.4), and 74.8% were female (n = 5,120). Most of the employees were white (56.8%), followed by Black (14.9%), Asian (10.7%), Hispanic (8.0%), and Other (9.0%). The majority of the employees were nurses (26.6%), followed by employees in operations (18.0%), administrative support (10.7%), research staff (8.7%), clinical technicians (7.3%), fellows/residents/interns (7.2%), clinical assistants (7.0%), and other (14.6%). 45.1% of the employees live within Hospital 1's service area, while 54.9% live within the service areas of the other 9 hospitals in the hospital network: 17.5% in Hospital 2; 10.7% in Hospital 3; 7.1% in Hospital 4; 6.7% in Hospital 5; 6.5% in Hospital 6; 3.9% in Hospital 7; 1.2% in Hospital 8; 0.9% in Hospital 9; 0.3% in Hospital 10. The total network has a mean COVID-19 census of 57.4 patients (SD = 61.3) and an average of 11.2 employees reporting symptoms each day (SD=15.1). While Hospital 1 has a mean COVID-19 census of 57.2 patients (SD = 61.5) and an average of 4.8 employees reporting symptoms each day (SD = 5.9), Hospital 10 has a mean COVID-19 census of 2.8 patients (SD = 3.5) and an average of 0.1 employees reporting symptoms each day (SD = 0.4) Employees filling attestations make up 0.8% of the total network's weighted service area (where service area population is weighted by the hospital's market share). Table 2 reports the descriptive statistics by hospital including the mean COVID-19 hospital census during the time of investigation and the average number of employees reporting symptoms per day.

Figure 2's upper panel plots the observed COVID-19 hospital census at Hospital 1 along with our predicted values and a 7-day forecast. In the lower panel, it plots the number of employees reporting symptoms who live in the service area of our main hospital each day.

Using the Dumitrescu and Hurlin (2012) test[6] for Granger non-causality in heterogeneous panel models on our full sample of 10 hospitals, we reject the null hypothesis that the number of employees reporting symptoms is not useful for forecasting future COVID-19 hospitalizations at any of the in-network hospitals (P < 0.001). We find that the optimal number of lags is 7 days, which minimizes the Bayesian information criteria.

We estimate a multivariable autoregressive linear regression model that includes each hospitals' daily COVID hospital census, and the number of employees reporting symptoms in each hospital's service area, with a lag of 7 days, to predict the daily COVID hospital census for each hospital over 209 days. The coefficient on Hospital 1's 7-day lag is 0.05 (95% CI: (0.02, 0.07); P < 0.001), which can be interpreted as meaning that twice as many employees reporting symptoms today is associated with a 5% increase COVID-19 hospitalizations in 7 days. At the mean for Hospital 1, this would correspond to 4.8 additional employees reporting symptoms today corresponds with 3 additional COVID-19 hospitalizations in 7 days. For Hospital 2, 7-day lag of symptoms was not statistically significantly different from zero (coefficient = -0.03; 95% CI: (-0.11,0.06); P < 0.001). For Hospital 3, a doubling of employees reporting symptoms today (1.3 additional reports at Hospital 3's mean) is associated with a 6% increase in COVID-19 hospitalizations in 7 days (95% CI: (0.00, 0.12); P < 0.001). For Hospital 4, a doubling of employees reporting symptoms (0.6 additional reports at Hospital 4's mean) is associated with an 8% increase in COVID-19 hospitalizations in 7 days (95% CI: (0.01, 0.16); P < 0.001). For Hospital 5, a doubling of employees reporting symptoms (0.6 additional reports at Hospital 5's mean) is associated with an 8% increase in COVID-19 hospitalizations in 7 days (95% CI: (0.02, 0.14); P < 0.001). For Hospital 6, the 7 day lag of symptoms was not statistically significantly different from zero (coefficient = 0.0, 95% CI = (-0.06, 0.06); P < 0.001 ). For Hospital 7, the 7 day lag of symptoms was not statistically significantly different from zero (coefficient = -0.01, 95% CI = (-0.09, 0.06); P < 0.001). For Hospital 8, the 7 day lag of symptoms was not statistically significantly different from zero (coefficient = -0.05, 95% CI = (-0.18, 0.07); P < 0.001). For Hospital 9, the 7 day lag of symptoms was not statistically significantly different from zero (coefficient = 0.0, 95% CI = (-0.04, 0.03); P < 0.001). For Hospital 10, the 7 day lag of symptoms was not statistically significantly different from zero (coefficient = -0.6, 95% CI = (-0.15, 0.03); P < 0.001).

In Table 2, we report the mean absolute error (MAE) and weighted mean absolute percentage error (WMAPE) by hospital for forecasts from November 5 to November 11. The model had a mean absolute error of 6.9 patients for the network. This corresponds to a weighted mean absolute percentage error of 1.5%. The individual hospitals had an MAE that ranged from 0.9 to 4.5 patients (WMAPE ranged from 2.1% to 16.1%). Hospital 1's MAE was 3.8 patients (WMAPE = 2.7%), Hospital 2's MAE was 1.4 patients (WMAPE = 5.7%), Hospital 3's MAE was 3.0 patients (WMAPE =5.7%), Hospital 4's MAE was 3.7 patients (WMAPE = 4.3%), Hospital 5's MAE was 4.5 patients (WMAPE = 16.1%) Hospital 6's MAE was 1.5 patients (WMAPE = 7.3%), Hospital 7's MAE was 1.3 patients (WMAPE = 3.6%), Hospital 8's MAE was 0.9 patients (WMAPE = 0.1%), Hospital 9's MAE was 0.1 patients (WMAPE = 14.2%), and Hospital 10's MAE was 1.3 patients (WMAPE = 4.9%).

As a secondary analysis, we also tried to predict positive cases, not just hospitalizations. We do this because the lag time between symptom onset and covid-19 test positivity is shorter than symptoms onset and hospitalization. However, this method relies more heavily on adequate and accessible testing, which is why hospitalizations are our preferred outcome measure of interest. Further, the state reporting agency only reports weekly case numbers by town, and only began reporting those numbers at the end of April. Thus, this analysis is done on a reduced sample at the week-level. From a Granger non-causality test for panel data, we cannot reject the null hypothesis that employees reporting symptoms is not useful for predicting changes in positive cases in the employees' home community (P = 0.15)

**Discussion**

With inconsistent access to broader testing that has varied over the course of the COVID-19 pandemic, many hospitals have relied on employees to accurately identify themselves as having symptoms of COVID-19 infections. In this study, we found a secondary use of these data to predict future hospitalizations in 7 days in the hospitals which service the communities in which these employees lived.

There is considerable prior work in epidemiological modelling as well as forecasting. In traditional epidemiological modelling, the Susceptible-Exposed-Infected-Recovered (SEIR) is one of the most common approaches. In June, several prominent SEIR national models averaged an mean absolute percentage error of 32.6%.[8] Another common forecasting approach is to use data from earlier outbreaks, adapting another locations curve to the local area, as the University of Washington's Institute for Health Metrics and Evaluation (IHME) model had done, yielding the lowest MAPE during this time period at 20.2%.[8] However, these are national models that predict further (10 weeks) into the future and predict deaths due to COVID-19. Our model is very localized to a specific area, predicts only into the immediate future (1 week) and forecasts hospitalizations, not deaths. The dynamic, rapidly changing COVID-19 policies for social distancing are challenging for these alternative approaches as they are highly sensitive to any change in transmissibility. In both of these alternative approaches, modellers must estimate not only the effect of any of these policy changes on the transmissibility of COVID-19, but also the rigor of their enforcement and adherence by the community. On the other hand, our approach requires no knowledge or estimation of ongoing policies. We instead use an early signal of community spread, employee reported symptoms, to predict future hospitalizations. This restricts our predictive capability to only the near future (7 days) which may not be sufficient for all purposes, but is sufficient to activate initial hospital and network emergency surge protocols. For example, New York state has mandated hospitals have a surge and flex response plan to expand operational bed capacity by 50% within 7 days.[9]

In future work, we plan to investigate other non-traditional data sources used previously in syndromic surveillance that may also act as surrogates for community spread. For example, when one is symptomatic with other types of infectious diseases such as influenza, one may make purchases of items to manage those symptoms such as facial tissues, orange juice, and other over-the-counter products.[10] It is plausible that other over-the-counter remedies such as tylenol, aspirin, and vitamin D that may be associated with COVID-19 could also be used. We also plan to investigate alternative methodological approaches. For example, we currently allow hospitals to have individual coefficients, independent of other hospitals. The coefficients are unlikely to be exactly the same, as hospitalization rates are likely to differ across communities, reflecting the diversity of demographics and underlying risk factors in different communities. However, these coefficients are unlikely to be entirely independent, so some soft parameter sharing across hospitals may be warranted.

*Limitations*

Our study does have several notable limitations. First, our estimate of COVID-19 daily hospital census represents the number of confirmed cases at each hospital. Thus, this estimate relies on testing data. Second, our independent variable relies on self-reported symptoms of employees reporting on-site for work. This depends on employees diligently filling out the symptom reporting and being honest about their experienced symptoms. While every effort is made to ensure all employees reporting to work are submitting an employee attestation form before they arrive, some employees may be on-site without filling out the employee attestation form. Employees may inconsistently report symptoms based on day of week, access to sick leave, and individual level of concern, all which create different patterns of missingness and bias in the attestation data itself. However, our findings suggest that despite these biases, employee reported symptoms still provide an adequate signal.

**Conclusion**

In conclusion, we find that daily employee symptom reporting can be a useful tool for predicting future COVID-19 hospitalizations in hospital networks.

**Figure 1: Employee Symptom Attestation Tool**

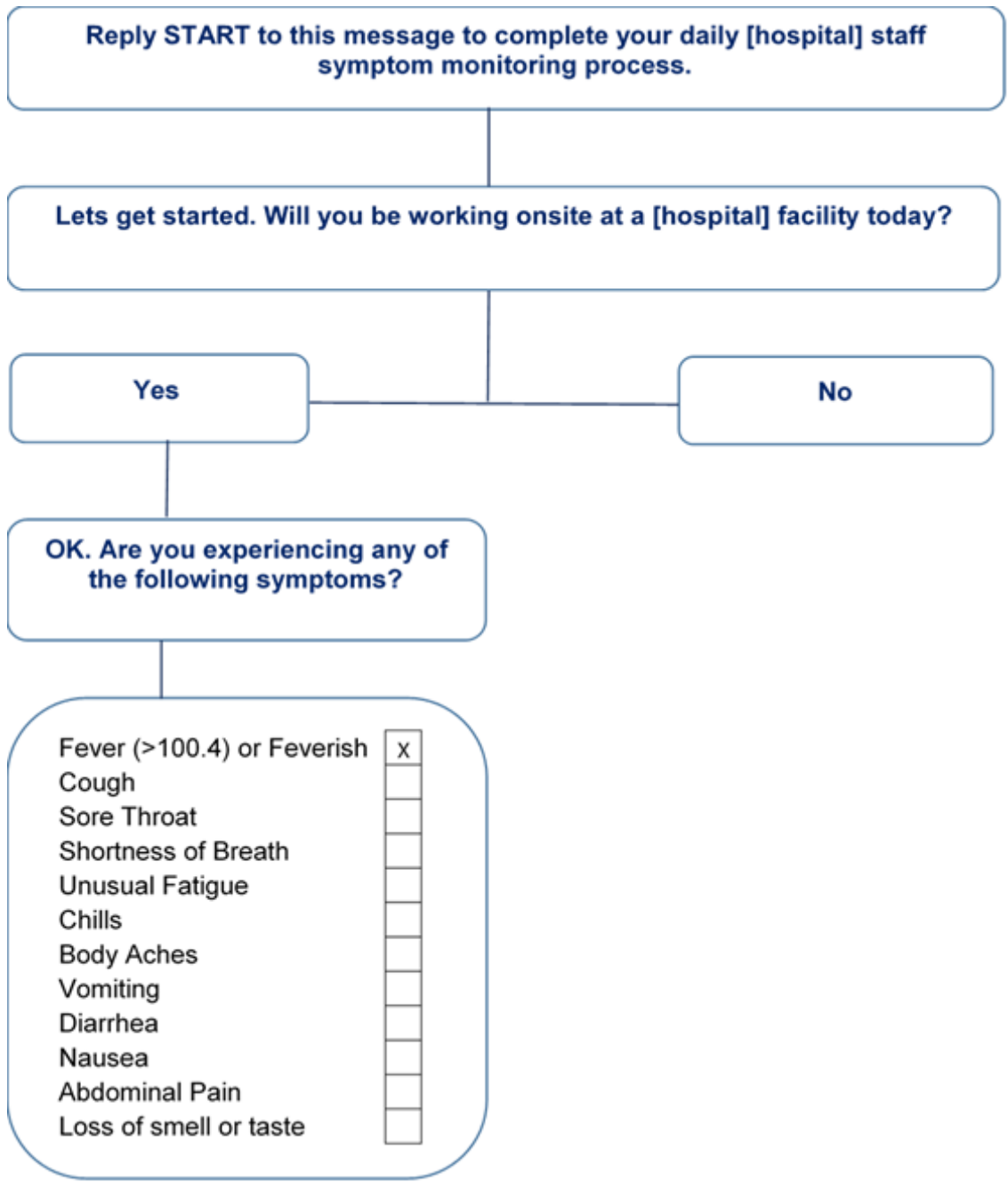

**Figure 2: Predicted Hospitalizations and Employee Symptom Reporting for Hospital 1**

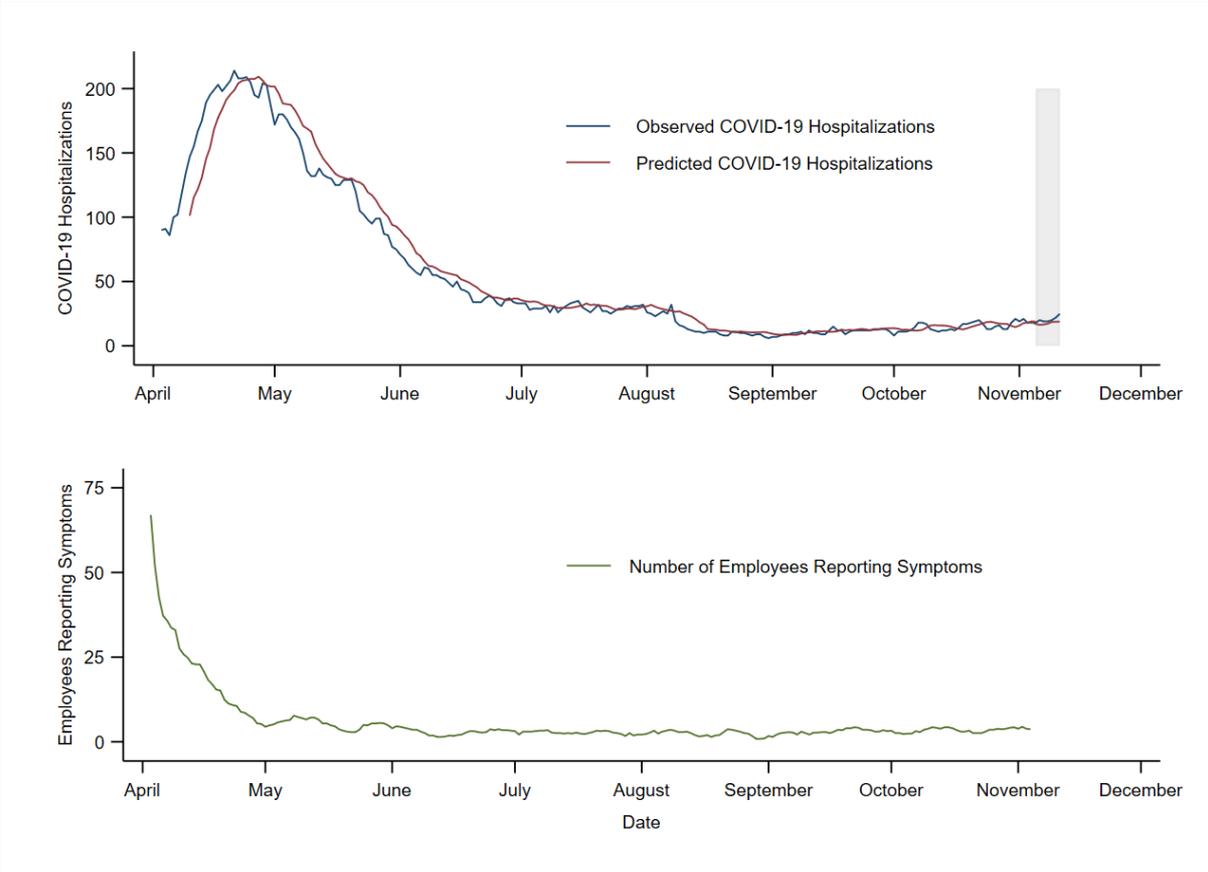

# Table 1: Descriptive Statistics of Employees

| | |
|---|---|
| Age, mean (SD) | 40.8 (13.6) |
| Years of Service, mean (SD) | 8.8 (10.4) |
| Female, n (%) | 5,120 (74.8%) |
| *Race* | |
|     White, n (%) | 3,884 (56.8%) |
|     Black, n (%) | 1,017 (14.9%) |
|     Asian, n (%) | 713 (10.4%) |
|     Hispanic, n (%) | 547 (8.0%) |
|     Other, n (%) | 680 (9.9%) |
| *Roles* | |
|     RN, n (%) | 1,818 (26.6%) |
|     Operations, n (%) | 1,230 (18.0%) |
|     Administrative Support, n (%) | 730 (10.7%) |
|     Research, n (%) | 597 (8.7%) |
|     Clinical Technician, n (%) | 501 (7.3%) |
|     Housestaff/Fellows/Residents/Interns, n (%) | 493 (7.2%) |
|     Assistant - Clinical, n (%) | 476 (7.0%) |
|     Other, n (%) | 996 (14.6%) |
| *Service Area Employee lives in (not where they work)* | |
|     Hospital 1, n (%) | 3,085 (45.1%) |
|     Hospital 2, n (%) | 1,197 (17.5%) |
|     Hospital 3, n (%) | 733 (10.7%) |
|     Hospital 4, n (%) | 488 (7.1%) |
|     Hospital 5, n (%) | 457 (6.7%) |
|     Hospital 6, n (%) | 444 (6.5%) |
|     Hospital 7, n (%) | 270 (3.9%) |
|     Hospital 8, n (%) | 81 (1.2%) |
|     Hospital 9, n (%) | 64 (0.9%) |
|     Hospital 10, n (%) | 22 (0.3%) |
| Total Employees | 6,841 |

**Table 2: Descriptive Statistics of Hospitals**

| Hospital | COVID-19 Hospitalizations Mean (SD) | Employees Reporting Symptoms Mean (SD) | Employees/Service Area Population[a] | MAE[b] (WMAPE)[c] |
|---|---|---|---|---|
| Hospital 1 | 57.2 (61.5) | 4.8 (5.9) | 0.8% | 3.8 (2.7%) |
| Hospital 2 | 10.3 (15.7) | 1.9 (2.7) | 4.6% | 1.4 (5.7%) |
| Hospital 3 | 8.1 (11.8) | 1.3 (2.1) | 1.6% | 3.0 (5.7%) |
| Hospital 4 | 27.8 (36.4) | 0.6 (1.3) | 0.4% | 3.7 (4.3%) |
| Hospital 5 | 13.0 (12.8) | 0.6 (1.1) | 0.5% | 4.5 (16.1%) |
| Hospital 6 | 4.0 (5.4) | 0.6 (1.2) | 2.0% | 1.5 (7.3%) |
| Hospital 7 | 15.8 (17.5) | 0.5 (1.0) | 0.3% | 1.3 (3.6%) |
| Hospital 8 | 8.6 (11.5) | 0.1 (0.4) | 0.3% | 0.9 (2.1%) |
| Hospital 9 | 0.0 (0.1) | 0.1 (0.4) | 2.7% | 0.1 (14.2%) |
| Hospital 10 | 2.8 (3.5) | 0.1 (0.4) | 0.2% | 1.3 (4.9%) |
| Total Network | 57.4 (61.3) | 11.2 (15.1) | 0.8% | 6.9 (1.5%) |

[a] Number of employees living in hospital's service area divided by the weighted service area's population, (weighted by the hospital's market share) displayed as a percentage.

[b] Mean Absolute Error

[c] Weighted Mean Absolute Percentage Error (mean absolute percentage error weighted by hospital census)

**References**

1. Saleem JJ, Read JM, Loehr BM, et al. Veterans' response to an automated text messaging protocol during the COVID-19 pandemic. *J Am Med Inform Assoc*. 2020;27(8):1300-1305.

2. Zhang H, Dimitrov D, Simpson L, et al. A Web-Based, Mobile-Responsive Application to Screen Health Care Workers for COVID-19 Symptoms: Rapid Design, Deployment, and Usage. *JMIR Form Res*. 2020;4(10):e19533.

3. Judson TJ, Odisho AY, Young JJ, et al. Implementation of a digital chatbot to screen health system employees during the COVID-19 pandemic. *J Am Med Inform Assoc*. 2020;27(9):1450-1455.

4. Nikolai LA, Meyer CG, Kremsner PG, Velavan TP. Asymptomatic SARS Coronavirus 2 infection: Invisible yet invincible. *Int J Infect Dis*. 2020;100:112-116.

5. Chou R, Dana T, Buckley DI, Selph S, Fu R, Totten AM. Epidemiology of and Risk Factors for Coronavirus Infection in Health Care Workers: A Living Rapid Review. *Ann Intern Med*. 2020;173(2):120-136.

6. Dumitrescu E-I, Hurlin C. Testing for Granger non-causality in heterogeneous panels. *Econ Model*. 2012;29(4):1450-1460.

7. Lampos V, Majumder MS, Yom-Tov E, et al. Tracking COVID-19 using online search. *arXiv [csSI]*. Published online March 18, 2020. http://arxiv.org/abs/2003.08086v10

8. IHME COVID-19 Forecasting Team. Modeling COVID-19 scenarios for the United States. *Nat Med*. Published online October 23, 2020. doi:10.1038/s41591-020-1132-9

9. Emergency Regulations. Accessed December 1, 2020. https://regs.health.ny.gov/regulations/emergency

10. Mandl KD, Overhage JM, Wagner MM, et al. Implementing syndromic surveillance: a practical guide informed by the early experience. *J Am Med Inform Assoc*. 2004;11(2):141-150.